\documentclass[preprint,floats,aps]{revtex4}
\usepackage{dcolumn}
\usepackage{bm}
\usepackage{graphicx}
\begin{document}
\topmargin=-0.3cm
\title{Kinetic Description of a Finite Temperature Meson Gas}
\author{ Zhi Guang Tan$^{1-2}$, Dai-Mei Zhou$^{1-2}$, S. Terranova$^1$ and A. Bonasera$^{1,3}$
 \footnote{Email: bonasera@lns.infn.it}}
\affiliation{
$^1$  Laboratorio Nazionale del Sud, Istituto Nazionale Di Fisica Nucleare,
      Via S. Sofia 44, I-95123 Catania, Italy \\
$^2$  Institute of Particle Physics, Huazhong Normal University, Wuhan,
      430079 China\\
$^3$ Libera Universita'  Kore, via della Cittadella 1, I-94100 Enna, Italy      
}
\begin{abstract}
A transport model  based on the mean free path approach for an interacting meson system at finite temperatures is discussed.  A transition to a quark gluon plasma  is included within the framework of the bag model.  We discuss some calculations for a pure meson gas where the Hagedorn limiting temperature is reproduced when including the experimentally observed resonances.  Next we include the
possibility for a QGP formation based on the MIT bag model.  The results obtained compare very well with Lattice QCD calculations. In particular the cross over to the QGP at about 175 MeV temperature  is nicely reproduced.
\end{abstract}
\maketitle

Numerical calculations of a hadron system and the transition to the quark-gluon plasma (QGP) at 
finite temperatures and zero baryonic density within a Lattice QCD (LQCD) approach are feasible nowadays  thanks to very performing computers\cite{karsch}.  Up to date LQCD results suggest that there is a cross over from a meson system to a QGP at a temperature of about 175 MeV \cite{karsch}.  These features could be experimentally confirmed or rejected  in 
relativistic heavy ion collisions (RHIC).  In particular high temperatures and very small baryonic densities can be obtained at the very high beam energies reached at the Brookhaven National Laboratory (BNL) and in the next future at the Large Hadron Collider (LHC) at CERN.  Because of the high complexity of
the problem, models are needed that take into account the relevant physics at such high energy densities, plus the possibility that the system is out of equilibrium during the collisions.  Of course microscopic, LQCD type, simulations for out of equilibrium-finite systems are out of reach at present.  On the other hand transport approaches  \cite{aich,gei}) have been very useful in the past in describing many features of lower energies heavy ion collisions.  Generalizations to relativistic energies of low energy heavy ion collisions \cite{cug,bert,aich} (known as  Boltzmann Uehling Uehlenbeck (BUU), Vlasov(VUU)/ Landau (LV)) have been proposed. They are based
 on  following the time evolution of each particle with a collision
occurring  if two of them come to a
closest distance less than or equal to $b_{max}=$ $\displaystyle{\sqrt
{\sigma_{tot}/\pi}}$. Here $\sigma_{tot}$ refers to the total cross section.   The method we discuss in
this work is known as Boltzmann Nordheim Vlasov (BNV) approach at low energies\cite{aldo,daimei}.  It
is based on the concept of the mean free path approach and we will discuss it in detail in the following
section.  In this paper we are interested in studying the equilibrium features of our model, i.e. the equation of state (EOS) effectively implemented in the model.
To this purpose, in our calculations we prepare two cluster systems of 300+300 pions which are initially boosted at a given energy and then collide similarly to heavy ion collisions.  This is done in order to study some non equilibrium effects as well.  The pions are enclosed in a box with periodic boundary conditions to simulate an infinite system.  The pions can collide elastically or inelastically according to the elementary cross sections which are parametrized from available data.  If we restrict our calculations to an hadron system only and we include all the measured resonances ($\rho$, $\omega$ etc..) we obtain the so called Hagedorn limiting temperature, i.e. when we increase the energy density of the system we do not obtain an increase of the temperature as well because  higher mass resonances are excited thus reducing the available kinetic energy.  We notice that, differently from the usually accepted Hagedorn resonance picture we obtain a limiting temperature also from the fact that some resonances (like the $\omega$) decay into 3 or more pions, thus part of the available kinetic energy is transformed into pion mass  lowering the temperature of the system.  Since those resonances are rather narrow it takes a long time before some $\pi$ $\pi$ collision forms exactly an $\omega$.

  We can easily include the possibility of a QGP using the bag model \cite{las,wong}. In fact, for each elementary hadron-hadron collision, we can calculate the local energy density and the pressure.  If such quantities overcome the Bag pressure and energy density then $n_q=n_{\bar q}$ quarks and antiquarks and $n_g$ gluons are created.  The number of quarks and gluons are calculated assuming local thermal equilibrium.  In this way we can simulate a hadron gas and its transition to the QGP.  Our results for different number of flavors $N_f$  compare very well to LQCD results.

 Quantum statistics (i.e. Pauli and Bose statistics) are  included
 similarly to   \cite{bert1}
 for Bose and \cite{sa04} for Fermi statistics. 

 \section{Formalism}

We follow the mean free path method as discussed
in\cite{aldo,daimei} and modified to include relativistic effects.  Briefly, for each event, at each time
 step dt and for each particle $i$ we search
for the closest one $j(i)$ in phase space i.e. we define the quantity
 \begin{equation}
\Xi_{ij}=\frac{r_{ij}}{v_{ij}dt}
\end{equation}

where ${r_{ij}}, {v_{ij}}$ and dt are the relative distances, velocities (relativistic) and $dt$ is
the time step used in the calculations.
Define a collision probability as:
 \begin{equation}
\Pi=\frac{v_{ij}dt}{\lambda}=\sigma\Pi_k(1\pm f_k) \rho(r_i)v_{ij}dt ,
\end{equation}
where $\rho(r_i)$ is the local density calculated at each time step. $f_k$ are the occupation functions of particle k after the collision occurs and the + (-)sign holds for bosons (fermions). The product runs over all the particles produced after the collisions.
  Note that the quantities
defined above are scalars and it is quite easy to show that they are
 Lorentz invariant.  For instance
in eq. (1) we have a distance  (${r_{ij}}$)  divided by a distance (${v_{ij}dt}$).

The physical meaning of the equation (1) is simple.
  We search for particles that are close
in coordinate space  and far away in velocity space, i.e.
particles with opposite momenta.  For instance at RHIC energies we
have a relative velocity of the order of the speed of light c for
particles of the target (T) and projectile (P)  respectively and
zero for both particles belonging to T or (P).  The latter
particles are automatically excluded from the first condition.
Once the two closest particles have been found, we calculate the
local density knowing the relative distance and the number of
particles near the colliding couple. From the relative energy and
the particles type we know the elementary cross sections and thus
the probability eq.(2).  A random number is taken and if smaller
than the calculated probability, a collision occurs.
 At the energy densities discussed in this work
  we parametrize the elementary cross sections according to the experimental data.  In particular we include all resonances and their decays up to about 2 GeV relative energies.  When including the possibility of a QGP as discussed below, to have higher energy resonances becomes unimportant since the QGP is formed already at smaller energies.

 Important differences for the collisions (thus for the dynamics) are due to the statistics.
 In fact,
at small energy densities we have essentially pions and $\rho$'s, while at higher energy densities  we can have quarks and gluons.
In the hadronic state essentially Bose
 Einstein statistics applies while in the partonic stage, both Bose and Fermi statistics apply.  Those features can be easily included in our approach through the occupation functions in eq.(2).
 
 The mean free path method discussed above has been studied in detail at low energies and it has been
shown to solve the Boltzmann eq. in the cases
 where an analytical solution is known \cite{aldo}.  Here we have
generalized the approach to keep into account relativistic effects.  The particles move on straight lines
during collisions since we have not implemented any force yet.  For short we name the method proposed
here as Relativistic Boltzman equation (ReB).  In order to test our approach, we first discuss some simple cases where analytical solutions are known to verify the sensitivity of our numerical solution.

 \section{An Ideal Pion Gas}

For illustration of our approach we first discuss a zero pion mass gas which interacts only through elastic collisions.  Because of these elastic collisions the system reaches a thermal equilibrium at temperature T.  In that limit various quantities such as energy densities and entropies versus temperature can be calculated both for a classical gas and a Bose gas.  In figure (1) we plot the energy density (top panel) and entropy (bottom) for 


\begin{figure}[ht]
\centerline{\hspace{-0.5in}
\includegraphics[width=4.5in,height=3.0in,angle=0]{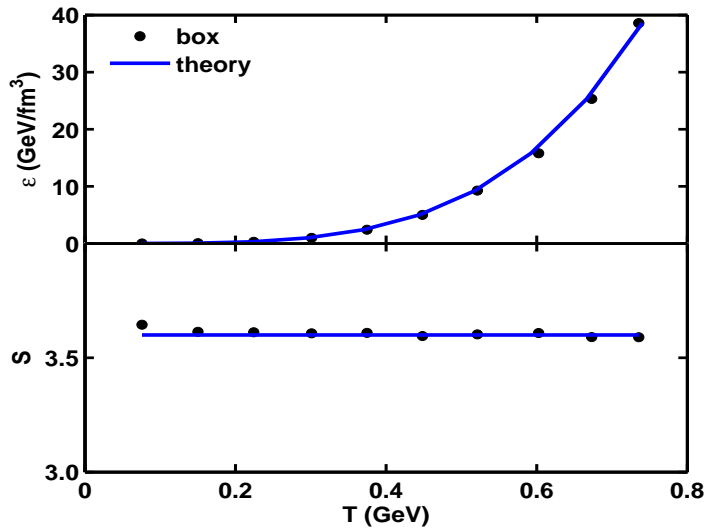}}
\vspace{0.2in} \caption{Energy density (top) and entropy (bottom) versus temperature for an ideal Bose-gas. The full lines represents the analytical, while the dots are our numerical result.  }
\label{fig1}
\end{figure}

 an ideal  Bose gas.  In this case in order to have zero chemical potential, at each initial energy corresponds a volume. The value of the entropy per particle of 3.6 is recovered \cite{las,wong}.  We stress that especially the entropy is very sensitive to the numerical approximation and to obtain such values we had to use a very small value of the time step and make an average over ensembles and time steps (since we are in equilibrium). 
 The energy density (bottom) and energy density divided $T^4$ for a gas of pions of mass 0.14GeV is displayed in figure(2).
 
\begin{figure}[ht]
\centerline{\hspace{-0.5in}
\includegraphics[width=4.5in,clip]{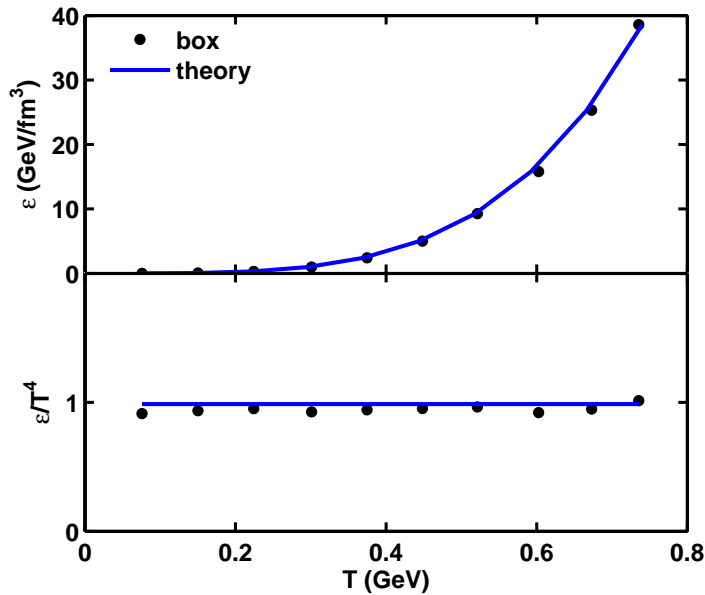}}
\vspace{0.2in} \caption{Energy density divided by $T^4$ (bottom) and energy density (top) versus temperature for a classical ideal gas of finite mass. The full lines represents the analytical, while the dots are our numerical result. }
\label{probx}
\end{figure}

\begin{figure}[ht]
\centerline{\hspace{-0.5in}
\includegraphics[width=4.5in,height=3.0in,angle=0]{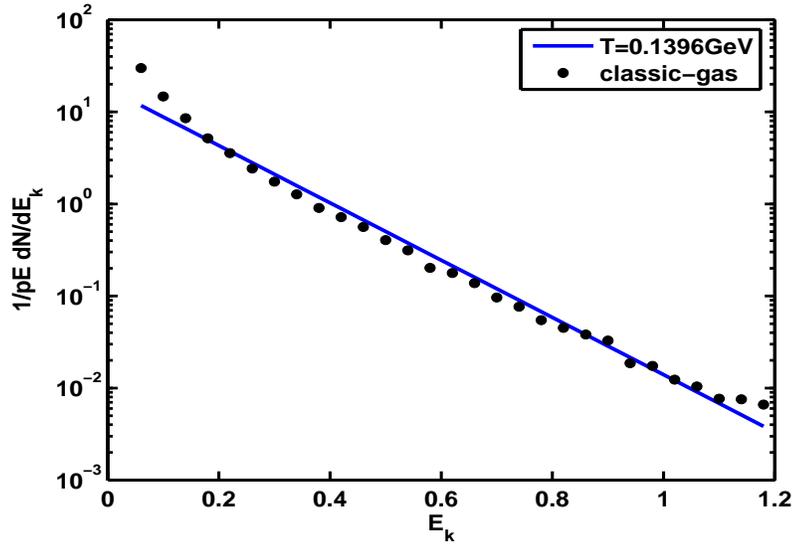}}
\vspace{0.2in} \caption{Particles kinetic energy distribution in the classical limit.  The fit is performed using a relativistic Maxwell-Boltzmann distribution.}
\label{aatau0}
\end{figure}

\begin{figure}[ht]
\centerline{\hspace{-0.5in}
\includegraphics[width=4.5in,height=3.0in,angle=0]{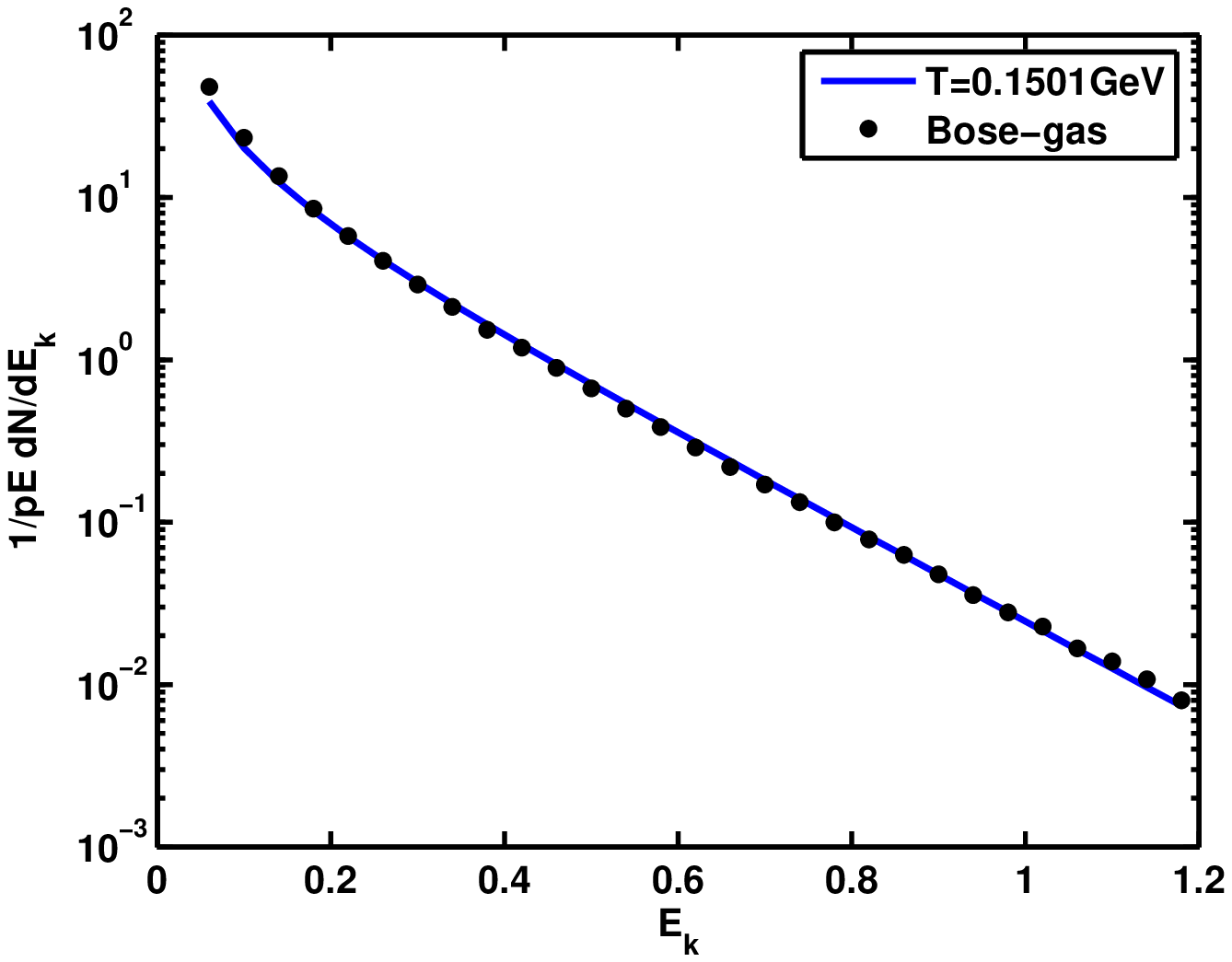}}
\vspace{0.2in} \caption{Same as figure(3) for a Bose gas.}
\label{fig4}
\end{figure}

The temperatures were determined in the calculations by fitting the numerical kinetic energy distributions with a relativistic Maxwell-Boltzmann distribution, fig.(3), or a Bose distribution in fig.(4)\cite{belk}.  We note in passing that especially for zero particle masses and high temperatures the difference between a classical and a Bose gas is negligible.

\section{From the Hagedorn limiting temperature to the QGP}

\begin{figure}[ht]
\centerline{\hspace{-0.5in}
\includegraphics[width=4.5in,height=3.0in,angle=0]{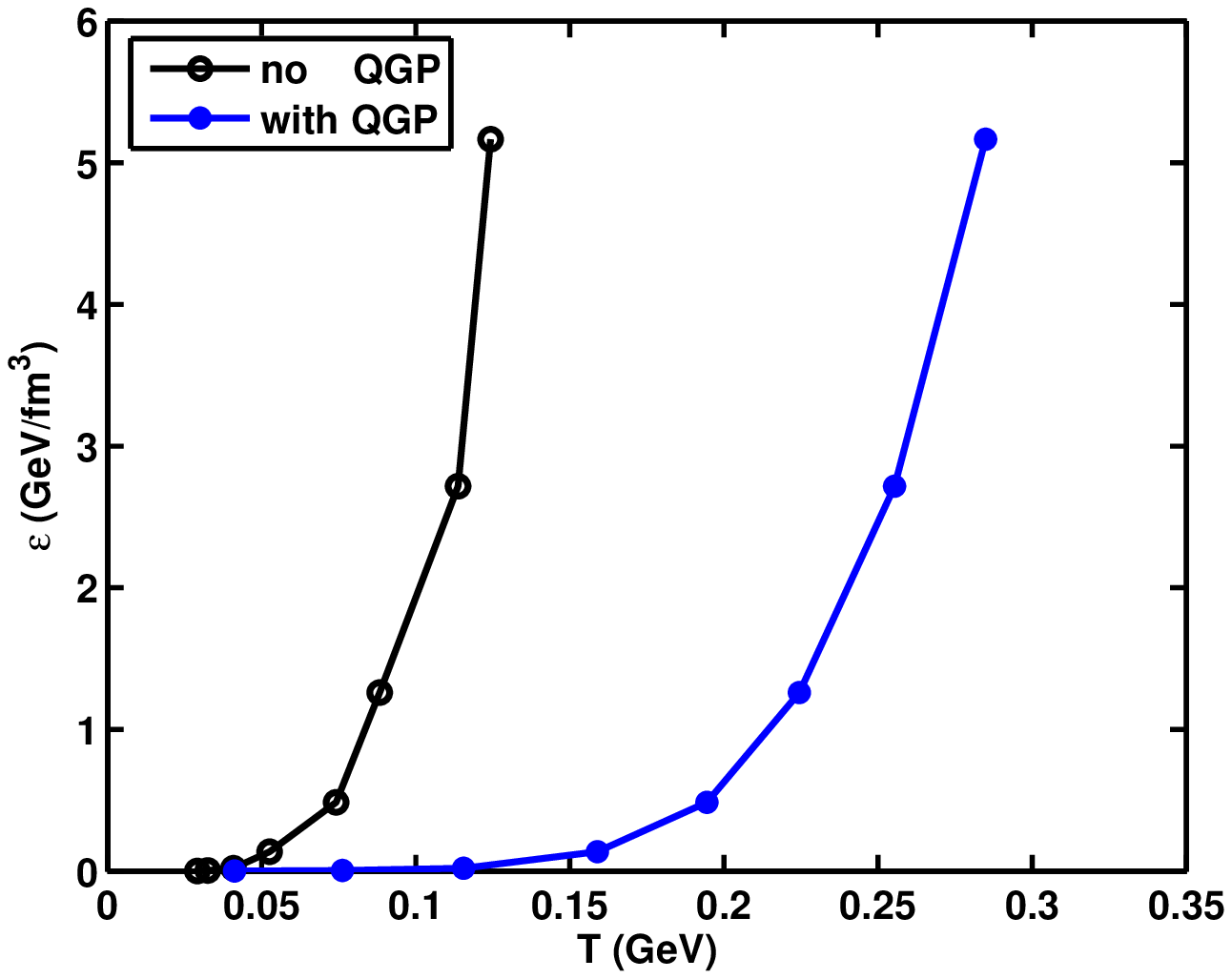}}
\vspace{0.2in} \caption{Energy density versus temperature for a purely meson gas with resonances (open circles) and for a QGP (full circles).  }
\label{fig5}
\end{figure}
In order to have a more realistic description of a meson gas we added the probabilities for resonances formation in  elementary meson-meson collisions and their decays.  The experimental data up to the $f_2(2340)$  resonance has been opportunely parametrized and channels for which there is no data available has been  included from theoretical models following\cite{bass}.  A resonance is followed in time until it decays according to its experimental lifetime.  The possibility of decays in 2 or more pions is included according to data.  In these preliminary calculations we have arbitrarily neglected the possibility of strange particles creation.  In this way we are restricting our results  to be able to compare to LQCD with 2 flavours.  A generalization to include strange particles is straightforward and we will discuss it in a following publication in order to explore the differences of calculations with and without strange particles.
\begin{figure}[ht]
\centerline{\hspace{-0.5in}
\includegraphics[width=5.5in,height=3.0in,angle=0]{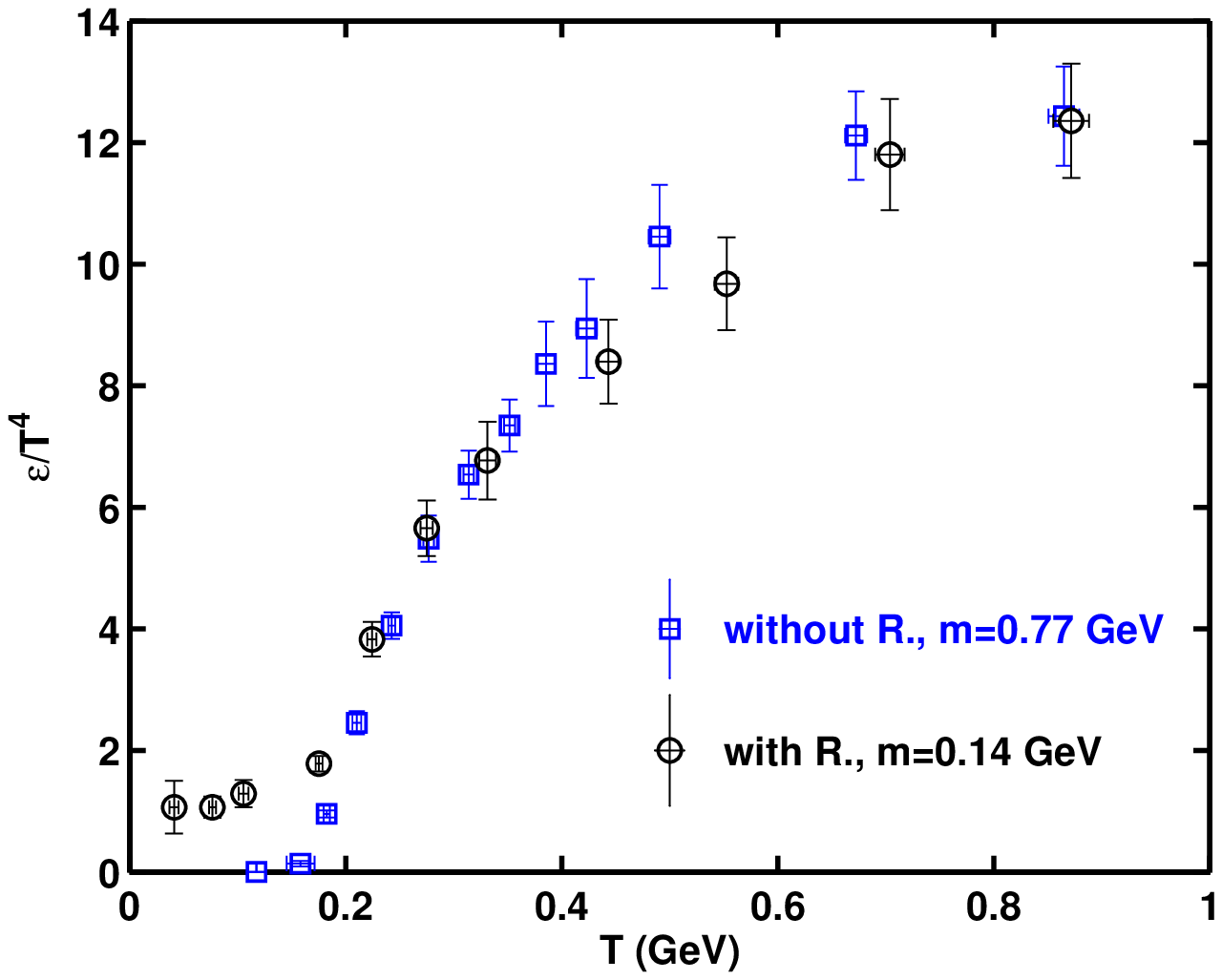}}
\vspace{0.2in} \caption{Equation of state including QGP formation and for pion masses equal to 0.14 GeV plus resonances  (open circles) and for a fixed meson mass of 0.77GeV (open squares).}
\label{fig6}
\end{figure}

In figure (5) the energy density versus temperature for the resonance meson gas is given by the open circles.  As it is shown in the figure there is a limiting temperature that the system can reach of about 140 MeV.  This is due to the fact that when increasing the energy density more and more resonances of higher masses are formed.  These resonances take kinetic energy away from the system thus lowering the temperature.  This is the Hagedorn scenario of the limiting temperature that can be reached in a purely hadronic system.  There is another reason why the system cannot overcome the Hagedorn temperature.  It is the possibility that a resonance such as the $\omega$ might decay into 3 $\pi$ thus converting kinetic energy into mass.  These resonances are however very narrow and to excite them is very difficult and it takes often a very long time.  However sooner or later they will be excited and more pions created.  Thus even if we would neglect higher resonances the limiting temperature will remain because of this possibility of decay into 3 or 4 pions.  This is of course a more visible effect if the possibility of decay of resonances into strange particles is included as well. 

We can now describe how to include a QGP in our approach.  First we recall that for a massless quark and gluon plasma in equilibrium the following relations hold for the pressure P, quark (antiquark and gluon) density $n_{q,(\bar q),(g)}$ and energy density $\epsilon$ versus temperature T\cite{las,wong,karsch}:

 \begin{equation}
P=g_{tot}\frac{\pi^2}{90}T^4;	
	n_q=n_{\bar q}=1.202\frac{3g_{q}}{4\pi^2} T^3;	
	n_g=1.202 \frac{g_g}{\pi^2}T^3;
\end{equation}

where $\epsilon=3P$, $g_{tot}=16+\frac{21}{2} N_f$, $N_f$ is the number of flavors.  In the MIT bag model\cite{las,wong} quarks and gluons are confined in the bag if the pressure is less than the critical pressure B that the bag can sustain.   Thus from the previous equation we can assume that the quarks and gluons are liberated in a collision if the energy density is larger then 3B.  This gives a critical energy density $\epsilon_c=3B=0.71 GeVfm^{-3}$ using $B^{1/4}=0.206 GeV$\cite{las,wong}.  For each h-h elementary collision we know  the volume and the energy of the  collision thus we can calculate the number of quarks, antiquarks and gluons liberated in the collision inverting equation (3).
We stress that these relations are 
 strictly valid in thermal equilibrium thus it is perfectly justified in this work since we are mainly discussing equilibrium features of our system.  The partons are followed in time exactly like the hadrons solving the transport equation.  In particular the partons can collide elastically with other partons and hadrons using a cross section of 1 mb  in eq.(1).  If in a collision between a parton and a hadron the local energy density is larger then the critical value, new partons are liberated from the hadron similarly to the mechanism discussed above.  The possibility of collisions among partons and hadrons is necessary in order that the total system can reach thermal equilibrium.

The results of our calculation when including the QGP are displayed in figs.(5) and (6).  In particular in figure (5) it is demonstrated the drastic difference with the pure hadrons calculations (full circle symbols) including the resonances.  In fact,  the Hagedorn limiting temperature is not present anymore, furthermore the energy density has a change of slope at about 0.17 GeV temperature.  The nature of this change is better explicated in figure(6) where the energy density divided the $T^4$ (see equation(3)) is plotted versus T.  The cases displayed are obtained by changing the pion mass and without including the possibility of resonance formation.  This is done in order to understand the role of the hadron masses. In fact, we find that increasing the mass gives a different critical temperature where the cross over is obtained.  This feature is easy to understand:  when calculating the (kinetic) energy density
in the elementary collisions which should overcome the critical value defined above, the corresponding 
 total energy density changes depending on the mass values for the colliding particles.

  In the figure the two main features of the system are seen, i.e. at low temperature the ratio 
$\frac{\epsilon}{T^4}$ is less than 1 which is the value expected for a mixture of bosons with masses, see fig.(2) \cite{las,wong}.  On the other hand, the ratio is  about 12 as it should be for a 2 flavors quark system larger than the value of about 8 obtained in LQCD calculations \cite{karsch}.  This is evidently due to the neglect of an interaction (attractive) among partons which is clearly important in LQCD.  This point will be discussed in more detail in a following work.

In our model  there is not a phase transition  but simply a crossover from an hadronic state   at low T to a state of partons at high T.  However, the crossover is rather visible and it should have some effect in experimental data at RHIC such as large fluctuations of D-mesons for instance \cite{terr05}.

The full calculations including hadronic resonances is displayed in figure (7) for $N_f=$ 2 (full circles) and $N_f=$0 (open squares).  For 2 flavors we obtain a cross over at about 175 MeV  temperature and a
higher temperature for the pure gluon case $N_f=$ 0.  These features are  at least in qualitative
agreement with microscopic LQCD results\cite{karsch} which strengthens our model somewhat. In fact, since the EOS of the two systems is similar, they should give similar consequences when features of heavy ion collisions are investigated.  We stress that our approach being a kinetic one could be easily extended and applied to relativistic heavy ion collisions which is the main purpose of this work.  Thus within this model we could study features of the collision with and without the phase transition and compare to data and/or be of guidance to more experimental investigations.

\begin{figure}[ht]
\centerline{\hspace{-0.5in}
\includegraphics[width=6.5in,height=4.0in,angle=0]{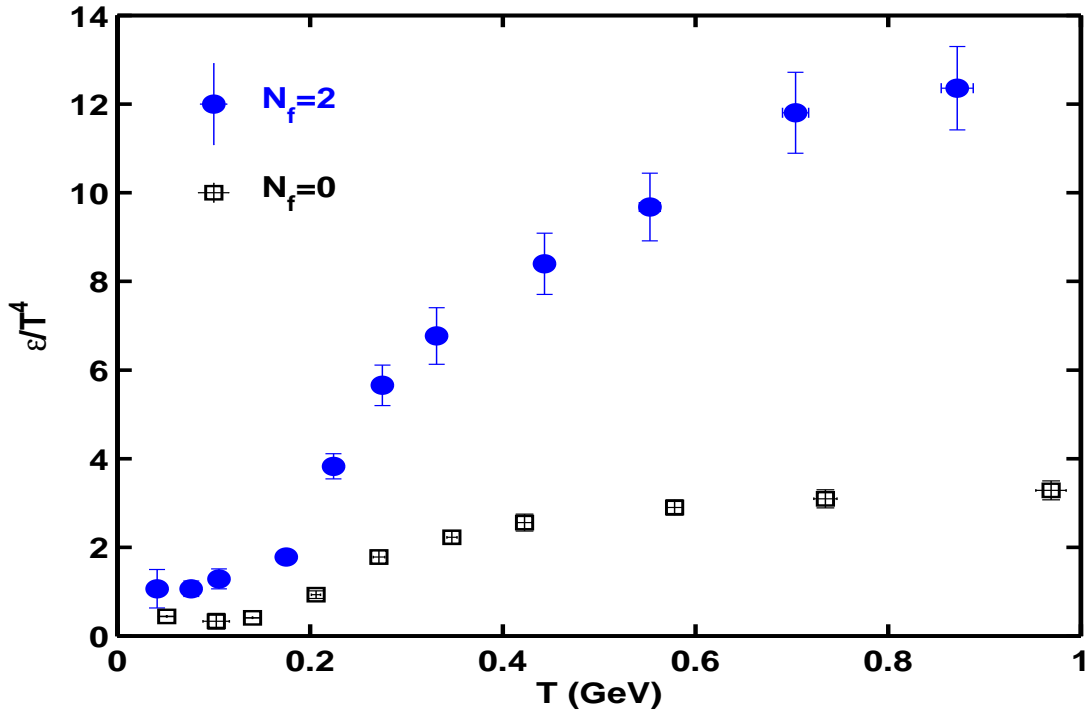}}
\vspace{0.2in} \caption{Energy density divided $T^4$ versus temperature for $N_f=2$ (full circles).  The LQCD results are given by the squares.}
\label{fig6}
\end{figure}
\section{conclusions}

In this work we have applied a recently introduced transport approach to
study the equation of state of a meson gas and its possible transitions to a Quark Gluon Plasma.
The model includes the possibility of resonances formation and their decay.  The possibility of a QGP
may be included based on the MIT bag model.  Quantum statistics can be easily included but we have 
seen that at the temperature discussed here those effects are negligible.  The Hagedorn limiting 
temperature is recovered for a pure hadronic system.  When including a QGP the obtained equation
of state is in qualitative agreement with LQCD calculations.  A better description could be obtained 
by introducing some interaction (attractive) among partons.  The results discussed here are obtained
in the case of zero and two flavors.  A generalization to the more realistic three flavors case is in 
progress.  The final goal of this work is to apply the model to relativistic heavy ion collisions at RHIC
and Cern energies. To this purpose we would like to have under control the ingredients of the model
such as the effectively implemented equation of state as well as a good reproduction of the available
data on elementary hadron-hadron collisions. Our efforts in these directions will be discussed in 
following papers.

Finally, Z.D.M and Z.G.T: acknowledge the financial support from  INFN and
Department of Phys. University of Catania in Italy (where most of
the work was performed) and NSFC (10347131) in China. This work is supported in part
 by the EU under contract CN/ASIA-LINK/008(094-791).  

\end{document}